\documentclass[10pt, prd, reprint, nofootinbib, superscriptaddress]{revtex4-1}

\usepackage[utf8]{inputenc}

\usepackage{amsmath}

\usepackage{amsfonts}
\usepackage{mathrsfs}

\usepackage{amssymb}
\usepackage{units}

\usepackage{graphicx}

\usepackage{placeins}

\usepackage{verbatim}

\usepackage{accents}

\usepackage{multirow}

\usepackage{color}

\usepackage{xpatch}

\newcommand{\mrm}{\mathrm}

\makeatletter

\let\old@dmathbeg\[
\let\old@dmathend\]

\newcommand{\rovnec}[1]{\old@dmathbeg#1\old@dmathend}
\newcommand{\rovcis}[2]{\begin{equation}#1\label{#2}\end{equation}}
\newcommand{\drovcis}[2]{\begin{equation}\begin{split}#1\end{split}\label{#2}\end{equation}}
\newcommand{\drovnec}[1]{\begin{equation*}\begin{split}#1\end{split}\end{equation*}} 
\newcommand{\provcis}[1]{\begin{align}#1\end{align}}
\newcommand{\provnec}[1]{\begin{align*}#1\end{align*}}

\newcommand{\rov}{\@ifstar\rovnec\rovcis}
\newcommand{\drov}{\@ifstar\drovnec\drovcis}
\newcommand{\prov}{\@ifstar\provnec\provcis}

\newcommand{\vast}{\bBigg@{4}}
\newcommand{\Vast}{\bBigg@{5}}

\patchcmd{\@ssect@ltx}
    {\addcontentsline{toc}{#1}{\protect\numberline{}#8}}
    {}
    {}
    {}

\makeatother

\DeclareMathOperator{\sgn}{sgn}

\DeclareMathOperator{\diffbold}{\mathbf{d}}
\newcommand{\bd}{\diffbold\!}

\newcommand{\msc}{\mathscr}
\newcommand{\mbs}{\boldsymbol}

\newcommand{\iDelta}{{\mit\Delta}}

\newcommand{\iSigma}{{\mit\Sigma}}

\DeclareFontEncoding{LGR}{}{}

\DeclareMathAlphabet{\msi}{OT1}{cmss}{m}{it}

\DeclareMathAlphabet{\mgr}{LGR}{cmr}{m}{n}

\newcommand{\equival}{\Longleftrightarrow}
\newcommand{\imply}{\Longrightarrow}
\DeclareMathOperator{\conj}{\&}

\renewcommand{\[}{\left[}
\renewcommand{\]}{\right]}

\newcommand{\f}{\!\left}
\newcommand{\ri}{\right} 

\newcommand{\der}[3][]{\frac{\mrm{d}^{#1}#2}{\mrm{d}{#3}^{#1}}}
\newcommand{\pder}[3][]{\frac{\partial^{#1}#2}{\partial{#3}^{#1}}}
\newcommand{\dpder}[2]{\pder[2]{#1}{#2}}

\newcommand{\nicepder}[3][]{\nicefrac{\partial^{#1}#2}{\partial{#3}^{#1}}}

\newcommand{\nicebpd}[2][]{\nicefrac{\boldsymbol\partial #1}{\boldsymbol\partial #2}}

\newcommand{\zrov}{{}\\{}}

\newcommand{\res}[2]{\left.#1\ri|_{#2}}
\newcommand{\lbl}{\label}
\newcommand{\rvt}{\ .}
\newcommand{\rvc}{\ ,}

\newcommand{\qt}[1]{``#1''}

\renewcommand{\(}{\left(}
\renewcommand{\)}{\right)}

\begin{document}

\title{Extraction of energy from an extremal rotating electrovacuum black hole:\\ Particle collisions along the axis of symmetry}

\author{F. Hejda}
\email{filip.hejda@tecnico.ulisboa.pt}
\affiliation{Centro Multidisciplinar de Astrofísica -- CENTRA, Departamento de
Física, Instituto Superior Técnico -- IST, Universidade de Lisboa -- UL, Avenida Rovisco Pais 1, 1049-001 Lisboa, Portugal}
\author{J. Bičák}
\email{bicak@mbox.troja.mff.cuni.cz}
\affiliation{Institute of Theoretical Physics, Faculty of Mathematics and Physics,
Charles University,
V Holešovičkách 2, 180\,00 Prague 8, Czech Republic}
\author{O. B. Zaslavskii}
\email{zaslav@ukr.net}
\affiliation{Department of Physics and Technology,
Kharkov V. N. Karazin National University,
4 Svoboda Square, Kharkov 61022, Ukraine
}
\affiliation{
Institute of Mathematics and Mechanics,
Kazan Federal University, 18 Kremlyovskaya Street,
Kazan 420008, Russia}

\begin{abstract}
High-energy collisions can occur for radially moving charged test particles in the extremal Reissner-Nordström spacetime if one of the particles is fine-tuned and the collision point is taken close to the horizon. This is an analogy of the Bañados-Silk-West (BSW) effect, first described for extremal Kerr black holes. However, it differs significantly in terms of energy extraction: unlike for the original BSW process, no unconditional upper bounds on mass and energy of an escaping test particle produced in the collision were found for the charged version. We show that these results can be replicated for the motion of charged test particles along the axis of a general extremal rotating electrovacuum black hole, also including the Schnittman-type process with reflected fine-tuned particles. 
This brings the possibility of high-energy extraction closer to astrophysical black holes, which can be fast spinning and have a small \qt{Wald  charge} due to interaction with external magnetic fields.
Nevertheless, we find numerous caveats that can make the energy extraction unfeasible despite the lack of unconditional kinematic bounds. 
\end{abstract}

\maketitle

\tableofcontents

\section{Introduction and conclusions}

The collisional Penrose process \cite{PirShK, PirSh} received a lot of attention after Bañados, Silk and West (BSW) \cite{BSW} described that for collisions involving fine-tuned (\qt{critical}) particles around an extremal Kerr black hole, the centre-of-mass collision energy grows without bound, if we take the collision point to the horizon radius. The original idea of a black hole acting as a particle supercollider to energies beyond the Planck scale turned out to be unfeasible due to numerous caveats (see e.g. \cite{BCGPS} for an early criticism and Section 3.1 in \cite{HK14} for summary and references). However, despite this, the \qt{BSW effect} is interesting in many ways. First, the BSW-type processes represent an intriguing theoretical issue on their own due to their ubiquity and their nature related to geometrical terms \cite{Zasl11b, Zasl11c}. Second, the investigation of the BSW-type phenomena led to the revision of upper bounds on extracted energy from a black hole, in particular when Schnittman \cite{Schnitt14} introduced a variant of the process with (nearly) critical particles reflected by the effective potential (see also \cite{BeBrCa}). Third, there is a variety of other collision processes applicable even to subextremal black holes; cf. for example \cite{GP11a,HK11a}. 
 (In addition, in \cite{TanatZasl13, TanatZasl14} it was discussed that under relatively weak conditions, processes with arbitrarily high collision energies can also survive deviations from geodeticity, e.g. due to backreaction.)
  The simpler effects for extremal black holes can be seen as the best-case scenario for those other processes. 

The original version of the BSW effect requires test particles with specific sign and magnitude of angular momentum; moreover, it is directly related to the Penrose process, which only works inside the ergosphere. 
However, an analogous effect is possible for radially moving charged particles in the extremal Reissner-Nordström spacetime \cite{Zasl11a}. Since this is a non-rotating charged black hole (without an ergosphere), the origin of the effect is purely electrostatic in this case. 
Furthermore, for this electrostatic variant it turned out that, in the test particle approximation, there is no bound on the mass and the energy of an escaping particle produced in the collision \cite{Zasl12c}. This is in sharp contrast to the original \qt{centrifugal} BSW effect, where unconditional bounds exist \cite{HaNeMi} (cf. also \cite{BPAH, Zasl12b}).

Astrophysical black holes are expected to be surrounded by external magnetic fields, and it has been proven in various contexts \cite{Wald74, ErnstWild} that a black hole immersed in an external magnetic field can maintain a non-zero charge. However, this so-called \qt{Wald charge} will be very small, and thus there is a good motivation to study generalisations of the radial electrostatic BSW-type effect to black holes with a smaller charge than in the extremal Reissner-Nordström case.

There are two natural ways of generalisation. 
First, one can include effects of angular momentum of the particles and of the dragging from the rotation of the black hole and study overlapping and transition between the electrostatic and the centrifugal BSW-type effect. A detailed analysis of this way of generalisation was given in \cite{a2} (concerning only the approach phase of the process). On the other hand, one can keep the restriction to \qt{purely radial} motion, which is  possible in any axially symmetric spacetime for particles moving along the axis of symmetry. 
In the present paper, we study this case and show that the interesting results for the extremal Reissner-Nordström black hole can be replicated even in models closer to astrophysical situations.

The paper is organised as follows. In Section \ref{sek:axmc}, we review the basic features of electrogeodesic motion along the axis of symmetry of a general stationary axially symmetric black-hole spacetime, including  the local definition of the critical particles. We review why they cannot approach the horizon for subextremal black holes 
and how they cause the divergent behaviour of the centre-of-mass energy in the limit of the collision point approaching the horizon radius. In \ref{odd:relax} we recall that the trajectory of a critical particle is approximated by an exponential relaxation towards the horizon radius. Because of this, any collision event involving a critical particle must always happen at a radius greater than the horizon radius. Therefore it makes sense to consider also particles that behave approximately as critical at a given collision radius 
(so-called nearly critical particles). In \ref{odd:clII} we show that doubly fine-tuned critical particles with infinite relaxation time exhibit an inverse power-law behaviour and thus approach the horizon radius much more slowly. 

In Section \ref{sek:krs} we study restrictions on the values of energy and charge of critical particles in order for them to be able to approach the radius of the degenerate horizon. Our discussion is based, similarly to \cite{a2}, on derivatives of a certain effective potential. In the Appendix we show how this approach can be rigorously related to the expansion coefficients of the radial equation of motion (including the relaxation time). In \ref{odd:kn} we give particular results for the extremal Kerr-Newman spacetime, which show that for small values of the black hole charge the critical particles must be highly relativistic in order to be able to approach the horizon radius.

In Section \ref{sek:extra} we deal with the energy extraction. First, we briefly review how to rearrange the conservation laws to prove that, for a $2\to2$ process, a collision of a critical particle with an incoming \qt{usual} (i.e. not fine-tuned) particle necessarily leads to the production of a nearly critical particle and an incoming usual particle. Then we study whether the produced nearly critical particle can escape and extract energy. 
We find that there are two threshold values, one for mass and one for energy. If the nearly critical particle is produced below/above the mass threshold, it is initially outgoing/incoming. Below the energy threshold the particle must be produced with such a value of charge that the corresponding critical energy will be lower than the actual energy, whereas above the threshold the critical energy corresponding to the charge is above the actual energy. These results qualitatively agree with the special case \cite{Zasl12c}. Here we focus on comparing the BSW-type process (collision with an incoming critical particle) and the Schnittman process (collision with an outgoing, reflected critical particle). 
For instance, a particle that is initially incoming with energy above the critical energy will fall into a black hole. Therefore, a particle that is produced with mass above the threshold must have the energy also above the respective threshold in order to avoid this. In 
\ref{odd:cav} we show that this may not be generally possible by considering a toy model of interactions of microscopic particles (cf. the \qt{neutral mass} problem). 
However, this problem occurs only for the BSW-type kinematics. Thus, the Schnittman variant again fares better. The problem is actually related to other two caveats for microscopic particles spotted earlier \cite{Zasl12c, NeMiHaKo}. As the energy of (nearly) critical particles is proportional to their charge, the (nearly) critical microscopic particles need to be highly relativistic (i.e. the \qt{energy feeding} problem), and also the produced particle must have a higher charge than the initial one, which limits the efficiency. Finally, we show that the energy feeding problem for microscopic particles may be reduced by six orders of magnitude if we go from the maximal value of the black hole charge for the Reissner-Nordström solution to some minimal value required for the processes to be possible. Despite this, the critical microscopic particles would still have to be highly relativistic, which is in sharp contrast to the behaviour for a small black hole charge (\qt{mega-BSW effect}) seen for the equatorial electrogeodesic case in \cite{a2}.
Concluding, our analysis of those details motivates further study of energy extraction from black holes through the generalised collisional Penrose process with charged particles.

\section{Motion and collisions of test particles along the axis}

\lbl{sek:axmc}

We start from a general axially symmetric stationary metric in the form
\rov{\mbs g=-N^2\bd t^2+g_{\varphi\varphi}\(\bd\varphi-\omega\bd t\)^2+g_{rr}\bd r^2+g_{\vartheta\vartheta}\bd\vartheta^2\rvt}{axst}
The metric components $g_{\varphi\varphi}, g_{rr}, g_{\vartheta\vartheta}$ and functions $N,\omega$ are independent of $t$ and $\varphi$; the metric is suitable to describe an equilibrium state of a black hole. 
We consider also an electromagnetic field with potential in the form 
\rov{\mbs A=A_t\bd t+A_\varphi\bd\varphi\rvc}{}
with $A_t,A_\varphi$ independent of $t$ and $\varphi$.
We assume that the outer black-hole horizon (where $N=0$) corresponds to $r=r_+$. For extremal black holes, we denote the position of their degenerate horizon by $r=r_0$. 

\subsection{Equations of motion and effective potential}

Let us consider the motion of charged test particles along the axis of symmetry.
The (semi)axis forms a two-dimensional submanifold. We can use two integrals of motion therein, which are related to the Killing vector $\nicebpd{t}$ and to the normalisation of the momentum. The axial motion is thus fully integrable. The first-order equations of motion for a particle with rest mass $m$ and charge $q$ read
\prov{p^t&=\frac{E+qA_t}{N^2}\rvc&p^r&=\sigma\sqrt{\frac{1}{N^2g_{rr}}\[\(E+qA_t\)^2-m^2N^2\]}\rvt\lbl{ptrgen}}
Here $E$ has the interpretation of the energy of the particle and $\sigma=\pm1$ distinguishes the outward/inward radial motion. 

The motion can be forbidden in some intervals of $r$ due to the presence of the square root in the expression for $p^r$; we require $\(p^r\)^2>0$. Let us assume that the product $N^2g_{rr}$ (which is equal to the volume element at the axis) is finite and non-vanishing at the axis, even for $N\to0$. 
For photons, we put $m=q=0$, and their kinematics is thus described by only one parameter $E$. Their motion is allowed for any $E\neq0$. In order to have $p^t>0$, we restrict to $E>0$. 

On the other hand, the kinematics of massive particles is characterised by two parameters $\varepsilon\equiv\nicefrac{E}{m}$ and $\tilde q\equiv\nicefrac{q}{m}$ (specific energy and specific charge). Denoting
\rov{W=\(\varepsilon+\tilde qA_t\)^2-N^2\rvc}{weff}
the condition for the motion to be allowed can be stated as $W\geqslant0$. Furthermore, since $N^2\geqslant0$ outside of the black hole, we can prescribe the decomposition of $W$,
\rov{W=\(\varepsilon-V_+\)\(\varepsilon-V_-\)\rvc}{wveff}
in terms of $V_\pm$ that read 
\rov{V_\pm=-\tilde q A_t\pm N\rvt}{veffpm}
In order for $W$ to be non-negative, it must hold either $\varepsilon\geqslant V_+$ or $\varepsilon\leqslant V_-$. However, only the first variant is consistent with $p^t>0$. Thus, we define $V\equiv V_+$ and consider only $\varepsilon\geqslant V$ as the condition for the motion to be allowed.
$\varepsilon=V$ is the condition for a turning point.

\subsection{Critical particles and collision energy}

Conditions $\(p^r\)^2>0$ and $p^t>0$ noted above have further implications. Particles with $E+qA_t^\mrm{H}>0$ ($A_t^\mrm{H}$ denotes $A_t$ at $r_+$) can fall into the black hole. For photons, this is the sole option as they have $E>0$, $q=0$. Thus, unlike in the equatorial case (see e.g. \cite{HaNeMi,Schnitt14}), photons along the axis are not so interesting. Turning to massive, charged particles, there is also a possibility for $\varepsilon+\tilde qA_t^\mrm{H}<0$, which corresponds to particles that cannot get close to the black hole, so it is also uninteresting for a generalised BSW effect. However, we can consider massive, charged particles with $\varepsilon+\tilde qA_t^\mrm{H}=0$. These are on the verge between the previous cases, and hence they are usually called critical particles.\footnote{Some authors (see e.g. \cite{BSW}) define the critical particles in a different way, such that they are on the brink of being able to reach the black hole \emph{from infinity}. We follow the \emph{local} definiton (cf. \cite{Zasl10}), which is more general. Both notions become compatible for extremal, asymptotically flat black hole spacetimes.} (To complement, particles that are not critical are called usual in the literature.)
Critical particles appear to have a turning point at the horizon radius, as seen 
e.g. through the fact that their specific energy, $\varepsilon_\mrm{cr}$, is equal to the value of the effective potential at the horizon 
\rov{\varepsilon_\mrm{cr}=-\tilde q\res{A_t}{r=r_+}=\res{V}{r=r_+}\rvt}{critgen}
Nevertheless, their trajectories actually do not reach a turning point, which we discuss in the next section(s). Why are the critical particles interesting for collision processes close to the horizon?

The formula for centre-of-mass collision energy reads (see e.g. \cite{GP11a} for more details)
\rov{E^2_\mrm{CM}=m_1^2+m_2^2-2g_{\alpha\beta}p_{(1)}^\alpha p_{(2)}^\beta\rvt}{}
Plugging in the equations of axial motion \eqref{ptrgen}, we get
\begin{widetext}
\rov{E_\mrm{CM}^2=m_1^2+m_2^2+2\frac{\(E_1+q_1A_t\)\(E_2+q_2A_t\)}{N^2}-\sigma_1\sigma_2\frac{2}{N^2}\sqrt{\(E_1+q_1A_t\)^2-m_1^2N^2}\sqrt{\(E_2+q_2A_t\)^2-m_2^2N^2}\rvt}{ecmax}
In order to consider the $N\to0$ limit (i.e. the collision point arbitrarily close to the horizon radius) 
in the case of a collision involving a critical particle, we examine the expansion of $W$ around $r_+$ with $\varepsilon+\tilde qA_t^\mrm{H}=0$: 
\rov{W\doteq-\res{\pder{\(N^2\)}{r}}{r=r_+}\(r-r_+\)+\res{\[\tilde q^2\(\pder{A_t}{r}\)^2-\frac{1}{2}\dpder{\(N^2\)}{r}\]}{r=r_+}\(r-r_+\)^2+\dots}{wcr}
The first radial derivative of $N^2$ at the horizon is proportional to surface gravity of the horizon and is non-negative. Let us first consider a generic, subextremal black hole (with non-zero surface gravity). We see from \eqref{wcr} that for some $r$ sufficiently close to $r_+$,  expression $W$ will become negative due to the linear term and, therefore, critical particles cannot approach $r_+$ for subextremal black holes (note that $W$ appears under the square root in \eqref{ptrgen}; cf. \eqref{weff}). In order to consider collisions with (precisely) critical particles arbitrarily close to the horizon radius, we thus have to turn to extremal black holes (see e.g. \cite{Zasl10,GP10a,a2} for more detailed analysis). Then we can use $N^2=\(r-r_0\)^2\tilde N^2$, where $\tilde N^2$ can be (at least formally) defined as 
\rov{{\tilde N}^2\equiv\sum_{n=2}^\infty\frac{1}{n!}\pder[n]{\(N^2\)}{r}\(r-r_0\)^{n-2}\rvt}{}
Evaluating \eqref{ecmax} for a collision of a critical particle $1$ and a usual particle $2$, we
 find that the leading order behavior in the $r\to r_0$ limit is 
\rov{E_\mrm{CM}^2\approx\frac{2}{r-r_0}\res{\left\{\frac{E_2+q_2A_t}{{\tilde N}^2}\[q_1\pder{A_t}{r}\mp\sqrt{q_1^2\(\pder{A_t}{r}\)^2-m_1^2{\tilde N}^2}\]\ri\}}{r=r_0,\vartheta=0}\rvt}{ecmcrhor}
\end{widetext}
The $\mp$ sign corresponds to $\sigma_1\sigma_2=\pm1$. However, for the usual particle one should consider only $\sigma_2=-1$ (cf. \cite{Zasl15b} for detailed reasoning). With this restriction, the $\mp$ sign means just $\sigma_1=\mp1$. The scenario with incoming particle $1$ (upper sign, $\sigma_1=-1$) was first described by Bañados, Silk and West for the extremal Kerr case in \cite{BSW} and was generalised to charged particles in \cite{Zasl11a}.  
 The collision process with an outgoing critical particle ($\sigma_1=+1$) was introduced by Schnittman \cite{Schnitt14} in a numerical study focused again on uncharged particles in the extremal Kerr spacetime. (Analytical treatment of the Schnittman process was considered e.g. in \cite{OHM, Zasl16b}.)

\subsection{Motion towards $r_0$ and nearly critical particles}

\lbl{odd:relax}

We have seen that for critical particles the centre-of-mass collision energy with an usual particle diverges in the limit $r\to r_0$. However, the energy attainable in such a thought experiment is always finite, although unbounded, because critical particles are not able to reach $r_0$ in a finite proper time. To demonstrate this, let us expand the equation of radial motion \eqref{ptrgen} near the radius of the degenerate horizon, 
\rov{\frac{p^r}{m}\equiv\der{r}{\tau}\doteq-\(r-r_0\)\res{\sqrt{\frac{1}{2{\tilde N}^2\tilde g_{rr}}\pder[2]{W}{r}}}{r=r_0}+\dots}{}
We denoted
\rov{\tilde g_{rr}\equiv\frac{N^2g_{rr}}{{\tilde N}^2}\rvt}{}
Then the approximate solution valid for late proper times is an exponential \qt{relaxation} towards $r_0$
\prov{r&\doteq r_0\[1+\exp\(-\frac{\tau}{\tau_\mrm{relax}}\)\]+\dots\\\frac{1}{\tau_\mrm{relax}}&=\res{\sqrt{\frac{1}{2{\tilde N}^2\tilde g_{rr}}\pder[2]{W}{r}}}{r=r_0}\rvt\lbl{taurel}}

Since no critical particle can ever reach $r_0$, the collision with another particle can only happen at some radius $r_\mrm{C}>r_0$. Because of this, the difference between usual and critical particles gets blurred. Indeed, a usual particle with energy very close to the critical energy will effectively behave as a critical particle at some radius $r_\mrm{C}$ close to $r_0$ provided that
\rov{\left|1-\frac{\varepsilon}{\varepsilon_\mrm{cr}}\ri|\sim\(\frac{r_\mrm{C}}{r_0}-1\)\rvt}{nceps}
Such particles are called nearly critical.

Nearly critical particles with $\varepsilon<\varepsilon_\mrm{cr}$ cannot fall into the black hole, and they have a turning point at some radius smaller than $r_\mrm{C}$. Thus, it makes sense to consider also the outgoing nearly critical particles. Furthermore, if the turning point is much closer to $r_0$ than the desired collision point $r_\mrm{C}$ or, more precisely, if 
\rov{0<\(1-\frac{\varepsilon}{\varepsilon_\mrm{cr}}\)\ll\(\frac{r_\mrm{C}}{r_0}-1\)\rvc}{schnapp}
such \emph{outgoing} nearly critical particles effectively behave as precisely critical at $r_\mrm{C}$. This is the motivation behind including outgoing critical particles in the Schnittman process.

\subsection{Remarks on class II critical particles}

\lbl{odd:clII}

One of the least studied aspects of the BSW-like phenomena is what happens when relaxation time $\tau_\mrm{relax}$ in \eqref{taurel} is infinite, i.e when the leading order of the expansion of $W$ in $r-r_0$ for a critical particle is the third one instead of the second. The critical particles with this property are called the \qt{class II} critical particles by Harada and Kimura \cite{HK11b} (\qt{class I} standing for the generic critical particles with finite $\tau_\mrm{relax}$). 

For the class II critical particles the expansion of the radial equation of motion \eqref{ptrgen} at $r_0$ turns to
\rov{\der{r}{\tau}\doteq-\(r-r_0\)^\frac{3}{2}\res{\sqrt{\frac{1}{6{\tilde N}^2\tilde g_{rr}}\pder[3]{W}{r}}}{r=r_0}+\dots}{urclII}
This leads to the following approximate solution (which describes outgoing critical particles for 
$\tau\to-\infty$ and ingoing ones for $\tau\to\infty$):
\rov{r=r_0+\frac{1}{\tau^2}\res{\(\frac{24{\tilde N}^2\tilde g_{rr}}{\pder[3]{W}{r}}\)}{r=r_0}+\dots}{iplclII}
This type of trajectory was previously considered in the equatorial geodesic case in \cite{Zasl15a} (be aware of typographic errors in equation (91) therein).
Because of their inverse power-law behaviour, class II critical particles approach $r_0$ much more slowly than class I critical particles with their exponential approach.

\section{Kinematic restrictions}

\lbl{sek:krs}

\subsection{General formulae for critical particles}

Critical particles can, in principle, approach the horizon radius only for extremal black holes. Whether their motion towards $r_0$ is really allowed will depend on their values of charge $\tilde q$ as well as on the properties of a particular extremal black hole spacetime. 

One way to figure out the conditions for the approach to be allowed is to look at the expansions of the radial equation of motion. For class I critical particles the relaxation time $\tau_\mrm{relax}$ in \eqref{taurel} must be a real number, and for class II critical particles the square root on the right-hand side of \eqref{urclII} must also be real.

The other way is to consider the $\varepsilon\geqslant V$ condition. Let us recall that the energy $\varepsilon_\mrm{cr}$ of a critical particle is equal to the value of $V$ at $r_0$ \eqref{critgen}.
Therefore, if the effective potential $V$ grows for $r>r_0$, we will get $\varepsilon_\mrm{cr}<V$, and the motion of the critical particle towards $r_0$ is forbidden. Thus, to see whether a critical particle can approach $r_0$, we need to check whether the first radial derivative of $V$ at $r_0$ is negative. Furthermore, we should also look at the second derivative of $V$ at $r_0$, since it will determine the trend of $V$, if the first one is zero.

However, both approaches are equivalent. For critical particles with $p^t>0$, it can be shown (see Appendix) that
\rov{\res{\sgn\dpder{W}{r}}{r=r_0}=-\res{\sgn\pder{V}{r}}{r=r_0}\rvt}{wveffddersignforw}
An analogous statement (cf. \eqref{clIIlogic}) can be made for class II critical particles,\footnote{Since equations \eqref{urclII}, \eqref{iplclII} and \eqref{clIIlogic} are applicable also to the equatorial case, they can be used to relate rigorously the results about the second derivative of $V$ in \cite{a2} (in Sections IV E and V B) to the kinematics of the class II critical particles.} and for our present setup, it actually holds that
\rov{\res{\pder[3]{W}{r}}{r=r_0,\vartheta=0}=-6\res{\(\tilde N\dpder{V}{r}\)}{r=r_0,\vartheta=0}\rvt}{wveffddderhorclII}

Let us proceed with the analysis based on $V$. For an extremal black hole, it is possible to write down an arbitrary ($n$-th) order derivative of $V$ with respect to $r$ as follows:
\rov{\pder[n]{V}{r}=-\tilde q\pder[n]{A_t}{r}+n\pder[n-1]{\tilde N}{r}+\(r-r_0\)\pder[n]{\tilde N}{r}\rvt}{}
At $r_0$, this simplifies to
\rov{\res{\pder[n]{V}{r}}{r=r_0}=\res{\(-\tilde q\pder[n]{A_t}{r}+n\pder[n-1]{\tilde N}{r}\)}{r=r_0,\vartheta=0}\rvt}{nderveffh}
It is possible to solve for the value of $\tilde q$, for which this expression becomes zero, and evaluate also the corresponding energy of the critical particle using \eqref{critgen}. In particular, for $n=1$, we get
\rov{\tilde q_\mrm{II}=\res{\frac{\tilde N}{\pder{A_t}{r}}}{r=r_0,\vartheta=0}\rvc}{qderveffzero}
and
\rov{\varepsilon_\mrm{II}=-\res{\frac{\tilde NA_t}{\pder{A_t}{r}}}{r=r_0,\vartheta=0}\rvt}{epsderveffzero}
If we denote
\rov{\alpha\equiv-\res{\frac{\pder{A_t}{r}}{\tilde NA_t}}{r=r_0,\vartheta=0}\rvc}{alphagen}
and assume $\alpha>0$ (this corresponds to a plausible choice of gauge constant for $A_t$), we can state that class I critical particles are allowed to approach $r_0$, whenever $\alpha\varepsilon>1$. For class II critical particles, it holds $\alpha\varepsilon=1$. Plugging \eqref{qderveffzero} into \eqref{nderveffh} with $n=2$, we obtain
\rov{\res{\dpder{V}{r}}{r=r_0}=\res{\(-\tilde N\frac{\dpder{A_t}{r}}{\pder{A_t}{r}}+2\pder{\tilde N}{r}\)}{r=r_0,\vartheta=0}\rvt}{dderveffhorclII}
Class II critical particles are allowed to approach $r_0$ if this expression is, for a given spacetime, negative.

Let us note that for a particle of any kind moving at a radius $r_\mrm{C}$ close to $r_0$, the expansion of $V$ to linear order can be expressed as
\rov{V=\varepsilon_\mrm{cr}+\tilde N_\mrm{H}\(1-\alpha\varepsilon_\mrm{cr}\)\(r_\mrm{C}-r_0\)+\dots}{veffnh}
Here $\varepsilon_\mrm{cr}\f(\tilde q\)$ is given by \eqref{critgen}; if $\alpha\varepsilon_\mrm{cr}\f(\tilde q\)>1$, the linear coefficient is negative. However, for particles that behave as nearly critical around $r_\mrm{C}$, their actual energy $\varepsilon$ is by definition \eqref{nceps} close to the critical one. Therefore, we can use $\alpha\varepsilon>1$ 
  also as a condition for the existence of escape trajectories of nearly critical particles (unless $\alpha\varepsilon$ is very close to 1), which is discussed in \ref{odd:krg}. 

\subsection{Results for the Kerr-Newman solution}

\lbl{odd:kn}

For the Kerr-Newman solution with mass $M$, angular momentum $aM$ (convention $a\geqslant0$), and charge $Q$ the metric \eqref{axst} reads
%\rov{\mbs g=-\frac{\iDelta\iSigma}{\msc A}\bd t^2+\frac{\msc A}{\iSigma}\sin^2\vartheta\[\bd\varphi-\frac{a}{\msc A}\(2Mr-Q^2\)\bd t\]^2+\frac{\iSigma}{\iDelta}\bd r^2+\iSigma\bd\vartheta^2\rvc}{kn}
\drov{\mbs g={}&{-\frac{\iDelta\iSigma}{\msc A}}\bd t^2+\frac{\msc A}{\iSigma}\sin^2\vartheta\[\bd\varphi-\frac{a}{\msc A}\(2Mr-Q^2\)\bd t\]^2+\zrov&+\frac{\iSigma}{\iDelta}\bd r^2+\iSigma\bd\vartheta^2\rvc}{kn}
where
%\prov{\iDelta&=r^2-2Mr+a^2+Q^2\rvc&\iSigma&=r^2+a^2\cos^2\vartheta\rvc&\msc A&=\(r^2+a^2\)^2-\iDelta a^2\sin^2\vartheta\rvt}
\prov{\iDelta&=r^2-2Mr+a^2+Q^2\rvc\nonumber\\\iSigma&=r^2+a^2\cos^2\vartheta\rvc\\\msc A&=\(r^2+a^2\)^2-\iDelta a^2\sin^2\vartheta\rvt\nonumber}
In the extremal case $M^2=Q^2+a^2$, so $\iDelta$ has a double root at $r_0\equiv M$. The electromagnetic potential is
\rov{\mbs A=-\frac{Qr}{\iSigma}\(\bd t-a\sin^2\vartheta\bd\varphi\)\rvt}{akn}

The effective potential for axial electrogeodesic motion (as given in (4.7) in \cite{BiStuBal}) reads
\rov{V=\frac{\tilde qQr}{r^2+a^2}+\sqrt{\frac{\iDelta}{r^2+a^2}}\rvt}{veffaxkn}
Let us note that for $a=0$, $Q^2=M^2$ and $\tilde q=\sgn Q$ we get $V\equiv1$. 

Particles moving along the axis of an extremal Kerr-Newman black hole are critical if their specific energy and charge are related by
\rov{\varepsilon_\mrm{cr}=\frac{\tilde qQ\sqrt{Q^2+a^2}}{Q^2+2a^2}\rvt}{critkn}

In general, the first radial derivative of $V$ at the degenerate horizon is
\rov{\res{\pder{V}{r}}{r=M}=-\tilde q\frac{Q^3}{\(Q^2+2a^2\)^2}+\frac{1}{\sqrt{Q^2+2a^2}}\rvt}{}
It becomes zero for particles with the specific charge given by 
\rov{\tilde q_\mrm{II}=\frac{\(Q^2+2a^2\)^\frac{3}{2}}{Q^3}\rvc}{qderveffzerokn}
and if these particles are critical, their specific energy is
\rov{\varepsilon_\mrm{II}\equiv\frac{1}{\alpha}=\frac{\sqrt{\(Q^2+a^2\)\(Q^2+2a^2\)}}{Q^2}\rvt}{epsderveffzerokn}
Class I critical particles are allowed to approach $r=M$, whenever $\alpha\varepsilon>1$. For class II critical particles $\alpha\varepsilon=1$. Let us note that $\alpha\leqslant1$ for any $Q$ and $a$. Therefore the condition $\alpha\varepsilon>1$ implies $\varepsilon>1$ (i.e. $E>m$). No bound critical particles can approach $r=M$ along the axis of the extremal Kerr-Newman spacetime.\footnote{This differs from the equatorial case; see \cite{a2}.} Furthermore, we can see that $\alpha\sim Q^2$. Thus, for $Q$ very small, only highly relativistic critical particles ($\varepsilon\gg1$) can approach $r=M$ along the axis.\footnote{Similarly, looking at the boundary value for charge $\tilde q_\mrm{II}$ \eqref{qderveffzerokn}, we see that only critical particles with $\left|\tilde q\ri|>1$ can approach $r=M$. And due to $Q^{-3}$ dependence in \eqref{qderveffzerokn}, for $\left|Q\ri|\ll M$, only critical particles with $\left|\tilde q\ri|\gg\varepsilon\gg1$ can approach $r=M$.}

The second derivative of $V$ at $r=M$ is
\rov{\res{\dpder{V}{r}}{r=M}=2\tilde qQ\sqrt{Q^2+a^2}\frac{Q^2-2a^2}{\(Q^2+2a^2\)^3}-\frac{2\sqrt{Q^2+a^2}}{\(Q^2+2a^2\)^\frac{3}{2}}\rvt}{}
Inserting \eqref{qderveffzerokn}, or evaluating 
 \eqref{dderveffhorclII}, we get 
\rov{\res{\dpder{V}{r}}{r=M}=-\frac{4a^2\sqrt{Q^2+a^2}}{Q^2\(Q^2+2a^2\)^\frac{3}{2}}\rvt}{}
This quantity is negative for $a\neq0$, and it blows up for $Q\to0$. Thus, class II critical particles are allowed to approach $r=M$ along the axis, except for the cases of the extremal Kerr solution (where there are no critical particles moving along the axis whatsoever) and of the extremal Reissner-Nordström solution (where $V$ becomes constant for $\alpha\varepsilon=1$). 

\section{Energy extraction}

\lbl{sek:extra}

\subsection{Application of conservation laws}

Let us now explore, in a simple setup, the possibility of energy extraction from black holes either by a BSW-type process occurring between particles moving along the axis, or by its Schnittman variant. We shall consider a scenario in which a (nearly) critical particle $1$ collides with an incoming usual particle $2$ close to the horizon radius $r_0$, they interact, and two new particles, $3$ and $4$, are produced. 
We impose the conservation of charge,
\rov{q_1+q_2=q_3+q_4\rvc}{}
and the conservation of (both components of) momentum at the point of collision. The time component gives us the conservation of energy
\rov{E_1+E_2=E_3+E_4\rvt}{}
In order to make the best use of the conservation of radial momentum, we shall note that for usual particles near the horizon, the following combination of the momentum components cancels up to the first order in $r-r_0$: 
\rov{N^2p^t-\sigma N\sqrt{g_{rr}}p^r\sim\(r-r_0\)^2\rvc}{ustrcanc}
whereas with the opposite sign
\rov{N^2p^t+\sigma N\sqrt{g_{rr}}p^r\doteq2\(E+qA_t^\mrm{H}\)+\dots}{ustrsum}
 contributes to the zeroth order.
In contrast, for the critical particles, or particles that behave as nearly critical around a desired collision radius $r_\mrm{C}$, both expressions are of the first order in $r_\mrm{C}-r_0$,
\rov{N^2p^t\pm N\sqrt{g_{rr}}p^r\sim\(r_\mrm{C}-r_0\)\rvt}{crittrsum}
To account consistently (at each order) for the effect of a particle labeled $i$, 
  which is not precisely critical, yet nearly critical, we define a formal expansion:
\rov{E_\mrm{cr}-E_i=C_{(i, 1)}\(r_\mrm{C}-r_0\)+C_{(i, 2)}\(r_\mrm{C}-r_0\)^2+\dots}{ncc}

Now, let us sum the conservation laws for the time and radial components of the momenta as follows:
\begin{widetext}
\rov{N^2\(p^t_{(1)}+p^t_{(2)}\)+N\sqrt{g_{rr}}\(p^r_{(1)}+p^r_{(2)}\)=N^2\(p^t_{(3)}+p^t_{(4)}\)+N\sqrt{g_{rr}}\(p^r_{(3)}+p^r_{(4)}\)\rvt}{trsum}
Considering expansion of this formula in $r_\mrm{C}-r_0$ and using \eqref{ustrcanc},\eqref{ustrsum} and \eqref{crittrsum}, we reach a conclusion (analogously to \cite{HaNeMi, Zasl12c}) that collision between a (nearly) critical particle $1$ and an incoming ($\sigma_2=-1$) usual particle $2$ at a radius $r_\mrm{C}$ close to $r_0$ must necessarily lead to the production of an incoming\footnote{See also \cite{Zasl15b} for a more detailed discussion on why it is impossible to produce outgoing usual particles near the horizon.} usual particle, to be denoted $4$, and a nearly critical particle, which we will label as $3$. 

Then, the leading (first) order of \eqref{trsum}, divided by $\tilde N_\mrm{H}$, implies
\rov{\alpha E_1+\sigma_1\sqrt{\alpha^2E_1^2-m_1^2}=\alpha E_3-\tilde C_3+\sigma_3\sqrt{\(\alpha E_3-\tilde C_3\)^2-m_3^2}\rvt}{consfin}
\begin{figure}
\centering
\input{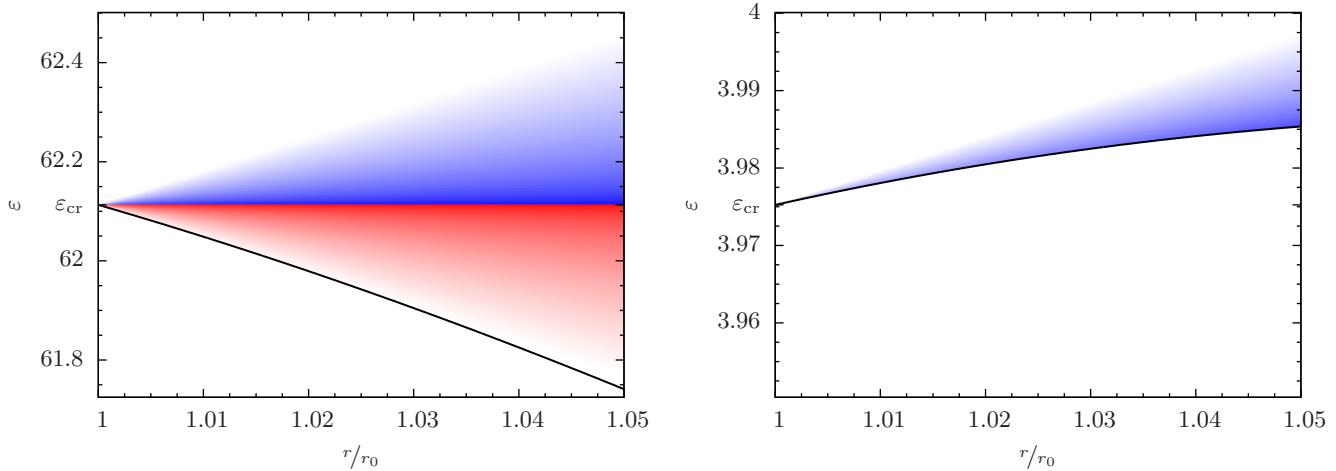}
\caption{Effective potential $V$ near the horizon radius $r_0=M$ of the extremal Kerr-Newman black hole with ${\nicefrac{Q}{M}=\nicefrac{1}{\sqrt{5}}}, {\nicefrac{a}{M}=\nicefrac{2}{\sqrt{5}}}$. The shading represents nearly critical particles; the varying height of the shaded areas illustrates the varying range of energies of particles that behave as nearly critical at a given $r$. The red color corresponds to \qt{$+$} ($\tilde C>0$), blue to \qt{$-$} ($\tilde C<0$). (See \eqref{ncc} and \eqref{ctil} for the definition of $\tilde C$.) {\bf Left:} For particles with $\tilde q=250$, $V$ is decreasing, and thus both signs of $\tilde C$ are allowed. {\bf Right:} For $\tilde q=16$, $V$ is increasing, and thus nearly critical particles can exist only with $\tilde C<0$. It can be seen that particles that behave as nearly critical at lower values of $r$ will get reflected by $V$ at some higher values of $r$, and thus they cannot escape.}
\lbl{fig:io}
\end{figure}
\end{widetext}
Here again $\sigma_1=-1$ corresponds to the BSW-type process, whereas $\sigma_1=+1$ to the Schnittman variant (and $\sigma_3=\pm1$ to outgoing/incoming particle $3$). Above we introduced
\rov{\tilde C_3\equiv\frac{C_{(3, 1)}}{\tilde N_\mrm{H}}\rvc}{ctil}
and, for simplification, we chose the particle $1$ precisely critical ($E_1=E_\mrm{cr}$, and hence $\tilde C_1=0$), which means that we are using the approximation \eqref{schnapp} for the Schnittman process. 

All the information about the spacetime coming into \eqref{consfin} is carried by the parameter $\alpha$ (defined in \eqref{alphagen}). Furthermore, if we denote the whole left-hand side of \eqref{consfin} as a new parameter\footnote{In order to keep the same letter for this quantity (introduced in \cite{HaNeMi} for the vacuum case, and followed e. g. by \cite{Zasl12b,Zasl12c}), we use a different font to distinguish it from the components of the electromagnetic potential.}
\rov{\msi{A}_1\equiv\alpha E_1+\sigma_1\sqrt{\alpha^2E_1^2-m_1^2}\rvc}{A1}
this parameter will express all the dependence on the properties of particle $1$.
Since we assumed $p^t>0$ and particle $1$ cannot be massless, we can make sure that $\msi{A}_1>0$.

Because we absorbed the difference between the BSW-type process and the Schnittman variant into the definition of the parameter $\msi{A}_1$, the discussion of kinematic regimes in the next section is the same for both. However, if we consider a particular model process, a significant distinction may appear, as we discuss in \ref{odd:cav}.

\subsection{Kinematic regimes}

\lbl{odd:krg}

Equation \eqref{consfin} enables us to determine whether and under which circumstances particle $3$ can escape and extract energy from a black hole. Particle $4$ necessarily falls into the black hole, which is the essence of a Penrose process. Let us note that particle $3$ can actually be produced in four different kinematic regimes, depending on the combination of the sign of $\tilde C_3$ and the sign variable $\sigma_3$. Following the classification in \cite{Zasl12c}, we will refer to the regimes with $C_3>0$ as \qt{$+$}, $C_3<0$ as \qt{$-$}, $\sigma_3=+1$ as \qt{OUT} and $\sigma_3=-1$ as \qt{IN}.

We analyse the different kinematic regimes from several points of view. First, we should understand which combinations are compatible with particle $3$ escaping from the vicinity of the black hole (see Figure \ref{fig:io} for illustration). For simplicity, let us assume a situation when effective potential $V$ for particle $3$ is well approximated by a linear function around $r_\mrm{C}$ (i.e. that $\alpha\varepsilon_3$ is not very close to $1$).

By definition, $\tilde C_3>0$ implies $\varepsilon_3<\varepsilon_\mrm{cr}$, and hence $\varepsilon_3<V$ (forbidden motion) at the horizon. Therefore, a particle produced with $\tilde C_3>0$ cannot fall into the black hole, and even if it is initially incoming, it must reach a turning point and turn to outgoing. Moreover, since it must hold that $\varepsilon_3>V$ at the radius $r_\mrm{C}$ where the particle is produced,  for such a particle the effective potential $V$ must be decreasing at $r_0$ (i.e. $\alpha\varepsilon_3>1$). Thus, we see that in kinematic regimes OUT$+$ and IN$+$ the local escape condition is satisfied automatically.

On the other hand, particle $3$ with $\tilde C_3<0$ will have $\varepsilon_3>V$ both at the horizon and at the point where it is produced (and also in between these points due to the assumption of $V$ being well approximated by a linear function).
 Thus, if a particle $3$ with $\tilde C_3<0$ is not produced as outgoing, it will fall into the black hole. Furthermore, if the effective potential $V$ is growing at $r_0$, i.e. $\alpha\varepsilon_3<1$,   particle $3$ will have a turning point at some radius greater than $r_\mrm{C}$ and it will not be able to escape even if it is produced as outgoing. Therefore, in the IN$-$ regime the escape is impossible and for OUT$-$ it depends on the trend of the effective potential $V$. (These findings are summarised in Table \ref{tab:reg}.)
 
Second, we should determine to what ranges of parameters of particle $3$ do the different kinematic regimes correspond. Then we can infer, whether the impossibility of escape in the IN$-$ regime leads to some bounds on parameters of the escaping particles, and more specifically, whether it does limit the efficiency of the collisional Penrose process, which is defined as 
\rov{\eta=\frac{E_3}{E_1+E_2}\rvt}{cpeff}
Solving \eqref{consfin} to express $\tilde C_3$ and $\sigma_3$, we get 
\prov{\tilde C_3&=\alpha E_3-\frac{1}{2}\(\msi{A}_1+\frac{m_3^2}{\msi{A}_1}\)\rvc\lbl{c3}\\
\sigma_3&=\sgn\f(\msi{A}_1-\frac{m_3^2}{\msi{A}_1}\)\equiv\sgn\f(\msi{A}_1-m_3\)\rvt}
From the second equation we see that the value of parameter $\msi{A}_1$ forms a threshold for $m_3$. If the interaction produces particle $3$ with a mass above the threshold, the particle must be incoming, if its mass is below the threshold, it must be outgoing.

Turning to parameter $\tilde C_3$ note that the solution \eqref{c3} satisfies the inequality
\rov{\tilde C_3\leqslant\alpha E_3-m_3\rvt}{c3Veff}
Therefore, if $\tilde C_3>0$, we must have $\alpha\varepsilon_3>1$, as we anticipated because particles with $\varepsilon<\varepsilon_\mrm{cr}$ can be produced only if effective potential $V$ is decreasing at $r_0$. (In general, one can see from \eqref{veffnh}, \eqref{ncc} and \eqref{ctil} that \eqref{c3Veff} is actually the linear order of the expansion in $r_\mrm{C}-r_0$ of the condition $\varepsilon\geqslant V$.)

Let us denote the value of $E_3$ for which $\tilde C_3=0$ as $\mu$: 
\rov{\mu=\frac{1}{2\alpha}\(\msi{A}_1+\frac{m_3^2}{\msi{A}_1}\)\rvt}{mu}
This quantity again 
represents  a threshold. 
If particle $3$ is produced with $E_3>\mu$, it must have such a value of charge 
that $E_\mrm{cr}\f(q_3\)>E_3$; if $E_3<\mu$, it must hold that $E_\mrm{cr}\f(q_3\)<E_3$. (Here $E_\mrm{cr}\f(q_3\)\equiv m_3\varepsilon_\mrm{cr}\f(q_3\)$; cf. \eqref{critgen}.)

\begin{table}
\caption{The four kinematic regimes for production of particle $3$.}
\begin{tabular}{r|c|c}
&$\sigma_3=+1$&$\sigma_3=-1$\\
\hline
\multirow{3}{*}{$\tilde C_3>0$}&OUT$+$&IN$+$\\
&$m_3<\msi{A}_1$, $E_3>\mu$&$m_3>\msi{A}_1$, $E_3>\mu$\\
&Guaranteed to escape&Guaranteed to escape\\
\hline
\multirow{3}{*}{$\tilde C_3\leqslant0$}&OUT$-$&IN$-$\\
&$m_3<\msi{A}_1$, $E_3\leqslant\mu$&$m_3>\msi{A}_1$, $E_3\leqslant\mu$\\
&Escapes if $\alpha E_3>m_3$&Falls inside the black hole
\end{tabular}
\lbl{tab:reg}
\end{table}

A summary of the results about the four kinematic regimes is given in Table \ref{tab:reg}. Let us note that 
these results resemble those for the special case of the Reissner-Nordström solution studied in \cite{Zasl12c}. In particular, there is still no unconditional upper bound on the energy or mass of particle $3$, in contrast with the geodesic (equatorial) case \cite{HaNeMi,Zasl12b}. (Such a possibility is often called the super-Penrose process.) However, the impossibility of escape in the IN$-$ regime means that whenever particle $3$ is produced with the mass above the threshold $\msi{A}_1$, its energy also must be above the threshold $\mu$ (which therefore acts as a lower bound on $E_3$ in this case). Conversely, whenever particle $3$ is produced with $E_3\leqslant\mu$, it must also have $m_3<\msi{A}_1$, otherwise it falls into the black hole. These requirements may not be compatible with the properties of a particular type of interaction that is responsible for producing particle $3$. This is the third aspect of the kinematic regimes that needs to be examined. In \ref{odd:cav} we consider a toy model, where this limitation gets highlighted (the \qt{neutral mass} problem).

Before carrying out this discussion, let us further note one interesting property of the OUT$-$ regime. Condition $m_3<\msi A_1$ (OUT) implies 
\rov{\alpha\mu<\msi A_1}{outmualpha}
 due to \eqref{mu}. From definition \eqref{A1}, we can derive an upper bound on $\msi A_1$. For the BSW-type process ($\sigma_1=-1$), we get $\msi A_1<\alpha E_1$, whereas for the Schnittman variant ($\sigma_1=1$), it is $\msi A_1<2\alpha E_1$. Combining with \eqref{outmualpha}, we get $\mu<E_1$ for the BSW-type effect and $\mu<2E_1$ for the Schnittman one. Using also the \qt{$-$} condition $E_3\leqslant\mu$, we get $E_3<E_1$ and $E_3<2E_1$, respectively. Therefore, we see that $E_3$ can never exceed $E_1$ in the OUT$-$ regime for the BSW-type process (preventing net energy extraction), whereas for the Schnittman variant $E_3>E_1$ is possible in this regime.

\subsection{Discussion of caveats}

\lbl{odd:cav}

\subsubsection{Energy feeding problem}

High efficiency $\eta$ of the collisional Penrose process means by definition \eqref{cpeff} that we can gain much more energy than we invest. However, despite a high value of $\eta$ the process may be \qt{inefficient} if the invested energy itself needs to be high in order for the process to occur. We  call this the \qt{energy feeding} problem. There are two different sources of this problem for particles moving along the axis. One of them was already mentioned in the discussion below equation \eqref{epsderveffzerokn}: for the extremal Kerr-Newman spacetime with a small value of charge ($\left|Q\ri|\ll M$), only highly relativistic critical particles can approach $r=M$ along the axis. This does not depend on the nature of the particles.

In contrast, the second source of the energy feeding problem comes into play only if we consider specifically processes involving microscopic particles that exhibit charge quantisation. For all those particles (known in nature) their specific charge $\left|\tilde q\ri|\gg1$. However, the specific energy of (nearly) critical particles is proportional to their specific charge (approximately) through relation \eqref{critgen}, or, in particular, by relation \eqref{critkn} for Kerr-Newman black holes. Therefore, such microscopic particles need to be highly relativistic ($\varepsilon\gg1$, i.e. $E\gg m$) in order to be (nearly) critical. Since the elementary charge is just one order of magnitude short of the Planck mass, they would actually have to be \emph{extremely} relativistic. This issue was previously noted in \cite{NeMiHaKo}, and it led the authors to introduce macroscopic objects acting as critical particles, which would make $\varepsilon\sim1$ possible (note $\varepsilon>1$ due to \eqref{epsderveffzerokn}).

\subsubsection{Neutral mass problem}

Although energy extraction by processes involving critical microscopic particles is already unfeasible due to the severe energy feeding problem, there are even further restrictions due to particle physics.
Since the energy of a (nearly) critical particle is proportional to its charge, we need $\left|q_3\ri|>\left|q_1\ri|$ in order to have $E_3>E_1$. For microscopic particles, this means that 
we need to turn to interactions involving atomic nuclei. (Let us note that 
such processes would actually \emph{not} benefit from high $E_\mrm{CM}$ due to a relatively low binding energy of nuclei, but here we focus on kinematic aspects.) One of the further problems was noted previously in \cite{Zasl12c}; stable nuclei have values of charge in a range that spans just two orders of magnitude. Thus, $E_3$ cannot exceed $E_1$ by more than a factor of $10^2$. However, the problems become much deeper, if we focus specifically on the BSW-type mechanism. The mass of stable nuclei generally increases faster than their charge due to an increasing share of neutrons (hence \qt{neutral mass}). Thus, for our model process with $q_3>q_1>0$ and $m_3>m_1$, it will also be more common than the opposite to have\footnote{Inequality \eqref{nm} could be the \qt{rule of thumb} even for macroscopic particles, as it is harder to hold together larger amounts of charge.}
\rov{\frac{q_3}{q_1}<\frac{m_3}{m_1}<\frac{m_3^2}{m_1^2}\rvt}{nm}
Now we should check whether this inequality is consistent with particle $3$ escaping. The problem again stems from the fact that critical microscopic particles are to be immensely relativistic. (At this point we exclude the possibility $Q\ll M$, i. e. $\alpha\ll 1$, which is revisited in \ref{po:scc}.)

Namely, for $E_1\gg m_1$ and $\sigma_1=-1$ parameter $\msi{A}_1$ \eqref{A1} will be very small; it can be approximated as
\rov{\msi{A}_1\doteq\frac{m_1^2}{2\alpha E_1}+\dots}{}
Given this, parameter $\mu$ \eqref{mu} gets large, and it is approximated as
\rov{\mu\approx E_1\frac{m_3^2}{m_1^2}+\dots}{muur}
Since certainly $\msi{A}_1<m_1$ and we assumed $m_1<m_3$, it will hold that $m_3>\msi{A}_1$. Thus, our nuclear reaction will occur in the IN regime. Condition $E_3>\mu$, which is required for the escape of particle $3$ in this regime (cf. Table \ref{tab:reg}), due to \eqref{muur} means
\rov{\frac{E_3}{E_1}>\frac{m_3^2}{m_1^2}\rvt}{}
As both energies are (approximately) proportional to the respective charges by the same factor, this translates to the relation
\rov{\frac{q_3}{q_1}>\frac{m_3^2}{m_1^2}\rvt}{nmesc}
However, this is the inequality opposite to \eqref{nm}. Therefore, we conclude that in our \qt{common nuclear process}, particle $3$ will be produced in the IN$-$ regime ($E_3<\mu$) and it will fall into the black hole.  Condition \eqref{nmesc} can be satisfied, e.g. with specific reactions with $q_3>q_1>0$, $m_3<m_1$, which are in principle also possible. Nevertheless, we see that there is a strong limitation on the BSW-type processes with microscopic particles.

However, if we turn to the Schnittman-type kinematics, the neutral mass problem is circumvented. In particular, for $E_1\gg m_1$ and $\sigma_1=+1$, parameter $\msi{A}_1$ is large, namely 
\rov{\msi{A}_1\approx2\alpha E_1+\dots}{}
Hence we infer $m_3<\msi{A}_1$ and particle $3$ to be produced in the OUT regime. Parameter $\mu$ will be large again, but this time dominated by the other term than before, i.e. 
\rov{\mu\approx E_1+\dots}{}
Since we assumed $q_3>q_1>0$, and hence $E_3>E_1$, particle $3$ will be produced in the OUT$+$ regime and will indeed escape. 

\subsubsection{Specific charge cutoff}

\lbl{po:scc}

The problems arising from the fact that critical microscopic particles have to be immensely relativistic can be reduced for the extremal Kerr-Newman solution if we consider $Q$ very small ($\left|Q\ri|\ll M$). However, we cannot decrease the required energy arbitrarily, because we run into the other source of the energy feeding problem, which is the proportionality $\alpha\sim Q^2$. Specific charges for all nuclei are roughly the same (of the same order), say $\tilde q_\mrm{nucl}$. Because of the critical condition \eqref{critkn}, all critical nuclei will also have values of specific energy of the same order. Thus, there will be a distinct transition.

Let us first consider a general value of $\tilde q$. Using \eqref{qderveffzerokn} we can define a value $\tilde Q_\mrm{min}$ of the specific charge of the black hole $\tilde Q\equiv\nicefrac{Q}{M}$, such that for $\tilde Q\sgn \tilde q<\tilde Q_\mrm{min}$ all the critical particles with the given value of $\tilde q$ would be forbidden to approach $r=M$. Using \eqref{critkn} or \eqref{epsderveffzerokn} we can also evaluate a corresponding specific energy $\varepsilon_\mrm{min}$. We obtain
\prov{\tilde Q_\mrm{min}&=\sqrt{\frac{2}{1+\left|\tilde q\ri|^\frac{2}{3}}}\rvc&\varepsilon_\mrm{min}&=\frac{\left|\tilde q\ri|^\frac{1}{3}}{\sqrt{2}}\sqrt{1+\left|\tilde q\ri|^\frac{2}{3}}\rvt}
However, for critical nuclei with $\tilde q_\mrm{nucl}\gg1$, we can use approximate expressions 
\prov{\tilde Q_\mrm{min}&\doteq\frac{\sqrt{2}}{\sqrt[3]{\tilde q_\mrm{nucl}}}\rvc&\varepsilon_\mrm{min}\approx\frac{\(\tilde q_\mrm{nucl}\)^\frac{2}{3}}{\sqrt{2}}\rvt}
Since $\tilde q_\mrm{nucl}$ is around $5\cdot10^{17}$, we get $\tilde Q_\mrm{min}$ of order $10^{-6}$ and $\varepsilon_\mrm{min}$ around $5\cdot10^{11}$. Therefore, for extremal Kerr-Newman black holes with $\tilde Q=\tilde Q_\mrm{min}$, the energy feeding problem for microscopic particles is reduced by six orders of magnitude as compared with the extremal Reissner-Nordström case (where $\varepsilon_\mrm{cr}=\tilde q_\mrm{nucl}$). Nevertheless, $\varepsilon_\mrm{min}\gg1$ in any case. Thus, we can never have non-relativistic critical microscopic particles approaching $r=M$ along the axis of an extremal Kerr-Newman black hole. This is very different from the \qt{mega-BSW} effect described in Section V E of \cite{a2} for equatorial charged critical particles.

\section*{Acknowledgements}

F. H. is grateful for the continued support of his supervisor at CENTRA, Professor J. P. S. Lemos.
F. H. thanks Fundação para a Ciência e a Tecnologia (Portugal) for funding his Ph.D. study at CENTRA through Grant No. PD/BD/113477/2015 awarded in the framework of the Doctoral Programme IDPASC-Portugal, and for travel support provided through UID/FIS/00099/2013 -- CENTRA. F. H. is also a proud member of COST Action CA16104 GWverse. J. B. acknowledges support from the Grant Agency of the Czech Republic, Grant No. GAČR 19-01850S. The work of O. Z. was performed according to the Russian Government Program of Competitive Growth of Kazan
Federal University.

\appendix

\section{Derivatives of $W$ and $V_\pm$}

Since the relation \eqref{wveff} among $W$ and $V_\pm$ is the same for both equatorial and axial motion, we can build on what was derived in \cite{a2} (in particular in the Appendix therein). Let us start with equation (34) of \cite{a2}, which states
\rov{\res{\pder[2]{W}{r}}{r=r_0,\varepsilon=\varepsilon_\mrm{cr}}=2\res{\(\pder{V_+}{r}\pder{V_-}{r}\)}{r=r_0}\rvt}{wveffdderhor}
Taking the third radial derivative of \eqref{wveff} above, we get
\begin{widetext}
\rov{\pder[3]{W}{r}=-\pder[3]{V_+}{r}\(\varepsilon-V_-\)+3\pder[2]{V_+}{r}\pder{V_-}{r}+3\pder{V_+}{r}\pder[2]{V_-}{r}-\(\varepsilon-V_+\)\pder[3]{V_-}{r}\rvt}{}
If we evaluate this relation for critical particles ($\varepsilon=\varepsilon_\mrm{cr}$) at the radius of the degenerate horizon (where $V_+=V_-=\varepsilon_\mrm{cr}$), it simplifies to
\rov{\res{\pder[3]{W}{r}}{r=r_0,\varepsilon=\varepsilon_\mrm{cr}}=3\res{\(\pder[2]{V_+}{r}\pder{V_-}{r}+\pder{V_+}{r}\pder[2]{V_-}{r}\)}{r=r_0}\rvt}{wveffddderhor}
Relations \eqref{wveffdderhor} and \eqref{wveffddderhor} have implications valid for both the equatorial and the axial motion, some of which can be further simplified in the axial case.

\subsubsection*{General case}

Because $V_+>V_-$ outside the horizon, though $V_+=V_-$ on the horizon, it must hold that
\rov{\res{\pder{V_+}{r}}{r=r_0}>\res{\pder{V_-}{r}}{r=r_0}\rvt}{derveffpmhor}
Using this with \eqref{wveffdderhor}, we arrive at the following two logical statements:
\prov{\res{\pder[2]{W}{r}}{r=r_0,\varepsilon=\varepsilon_\mrm{cr}}<0\equival\(\res{\pder{V_+}{r}}{r=r_0}>0\)\conj\(\res{\pder{V_-}{r}}{r=r_0}<0\)\rvc\\
\res{\pder[2]{W}{r}}{r=r_0,\varepsilon=\varepsilon_\mrm{cr}}>0\equival\(\res{\pder{V_+}{r}}{r=r_0}<0\)\textrm{or}\(\res{\pder{V_-}{r}}{r=r_0}>0\)\rvt}
It is easy to check that the two variants in the second statement correspond to the critical particle having $p^t>0$ or $p^t<0$, respectively.\footnote{Note that $p_t>0$ corresponds to $\varepsilon\geqslant V_+$ and $p_t<0$ to $\varepsilon\leqslant V_-$, and for critical particles their energy $\varepsilon_\mrm{cr}=\res{V_+}{r_0}=\res{V_-}{r_0}$. Thus $\res{\nicepder{V_+}{r}}{r_0}<0$ corresponds to critical particles with $p_t>0$ and $\res{\nicepder{V_-}{r}}{r_0}>0$ to those with $p_t<0$.} Thus, with restriction to $p^t>0$, equation \eqref{wveffddersignforw} follows. 

Using \eqref{derveffpmhor} also with \eqref{wveffddderhor}, we get a statement analogous to \eqref{wveffddersignforw} for class II critical particles:
\rov{\varepsilon=\varepsilon_\mrm{cr}\conj\res{\pder{V_+}{r}}{r=r_0}=0\imply\(\res{\dpder{W}{r}}{r=r_0}=0\)\conj\(\res{\sgn\pder[3]{W}{r}}{r=r_0}=-\res{\sgn\dpder{V_+}{r}}{r=r_0}\)\rvt}{clIIlogic}
\end{widetext}

\subsubsection*{Axial case}

For motion along the axis \eqref{clIIlogic} can be further refined. From 
 definition \eqref{veffpm}, we can calculate
\rov{\res{\pder{V_-}{r}}{r=r_0,\vartheta=0}=-\res{\(\tilde q\pder{A_t}{r}+\tilde N\)}{r=r_0,\vartheta=0}\rvt}{}
Using the value of $\tilde q$ for class II critical particles \eqref{qderveffzero}, we get 
\rov{\res{\pder{V_-}{r}}{r=r_0,\vartheta=0}=-2\res{\tilde N}{r=r_0,\vartheta=0}\rvc}{}
and if we plug the result into \eqref{wveffddderhor}, we arrive at \eqref{wveffddderhorclII}.

\end{document}